\documentclass{article}

\usepackage[english]{babel}

\usepackage[letterpaper,top=2cm,bottom=2cm,left=3cm,right=3cm,marginparwidth=1.75cm]{geometry}

\usepackage{amsmath}
\usepackage{graphicx}
\usepackage[colorlinks=true, allcolors=blue]{hyperref}

\usepackage{authblk}
\usepackage{mathrsfs}

\title{An exact solution for the magnetic diffusion problem with a step-function resistivity model}

\author[1]{Bo Xiao*}
\author[1]{Ganghua Wang}
\author[2]{Li Zhao}
\author[2]{Chunsheng Feng}
\author[2]{Shi Shu}
\affil[1]{Institute of Fluid Physics, China Academy of Engineering Physics, Mianyang 621999, P. R. China}
\affil[2]{School of Mathematics and Computational Science, Xiangtan University, Xiangtan 411105, P. R. China}
\affil[*]{homenature@139.com}

\date{}

\begin{document}
\maketitle

\begin{abstract}
In the magnetic diffusion problem, a magnetic diffusion equation is coupled by an Ohmic heating energy equation.
The Ohmic heating can make the magnetic diffusion coefficient (i. e., the resistivity) vary violently, and make the diffusion a highly nonlinear process.
For this reason, the problem is normally very hard to be solved analytically.
In this article, under the condition of a step-function resistivity and a constant boundary magnetic field, we successfully derived an exact solution for this nonlinear problem, 
which should be an interesting thing in the area of partial differential equations.
What's more, the solution could serve as a valuable benchmark example for testing simulation methods of the magnetic diffusion problem.
\end{abstract}

\section{Introduction}

In this paper, the so called ``magnetic diffusion problem'' is described by a magnetic diffusion equation and an Ohmic heating induced internal energy equation. 
The two equations are coupled as follows:
the magnetic diffusion evolves the magnetic field distribution, whose gradient determines the Ohmic heating rate; on the other hand, Ohmic heating changes the internal energy distribution, which affects the resistivity distribution through the $\eta-e$ relation (resistivity to internal energy density relation), and resistivity in return determines the magnetic diffusion rate. 
The $\eta-e$ relation plays a key role in determining the coupling strength of these two equations. For example, when the resistivity is a constant independent of internal energy, the magnetic diffusion equation becomes an independent one. On the contrary, if the resistivity depends strongly on internal energy, then the coupling of the two equations would become so strong that it makes the magnetic diffusion a highly nonlinear process.

This paper focus on the magnetic diffusion problem with a step-function resistivity model, in which the resistivity is a small constant $\eta_{\text{S}}$ when the internal energy density is below some critical point $e_{\text{c}}$, while it jumps to a value $\eta_{\text{L}}$ that is much larger than $\eta_{\text{S}}$ when the internal energy density is above $e_{\text{c}}$.
The step-function model of resistivity is not just a theoretical fancy, but a simplified description of metals' real resistances, with the jump of resistivity at $e_{\text{c}}$ representing the dramatic growth of resistivity around the gasification temperature of a metal \cite{burgess1986}.
Despite its simpleness in appearance, this step-function model of resistivity can make troubles to the numerical simulation of magnetic diffusion problem,
just considering that different regions have distinguished magnetic diffusion rates and boundaries between these regions vary violently with the diffusion.
It may make simulations hard to achieve convergent results, or lead to artificial instabilities around the magnetic field diffusion front in two or three dimensional simulations, 
according to our experiences.

The purpose of this paper is however not to construct numerical schemes for simulation of the magnetic diffusion problem.
Instead, It focuses on deriving an exact analytical solution for the problem, by assuming a step-function resistivity model and a constant magnetic field boundary condition.
An exact analytical solution, besides of interests in its own theoretical part, can support the simulation works in several aspects.
First it can be used as an idealized benchmark for testing simulation methods.
In addition, innovative simulation methods for the magnetic diffusion problem may be constructed by utilizing information from the analytical solution, like the ``shock-fitting'' methods \cite{zouDY1986} constructed for simulating compressible fluids with shock waves.

In literature's, works that can be found dedicated in studying analytical solutions for the magnetic diffusion problem are few.
One typical work is by Schnitzer \cite{schnitzer2014}, which derived analytical solutions for magnetic diffusion problem under the condition of a power-function resistivity model and a power-law rising surface magnetic field.
Some other works have been carried out by us \cite{xiao2016, yan2021}. In those works, \emph{approximated} analytical solutions for the magnetic diffusion problem under the condition of a step-function resistivity model and a constant magnetic field boundary condition were derived.
The current paper, on some aspects, can be viewed as an extended work from those earlier works, with the approximated solutions being improved to an \emph{exact} one.

The remaining parts of this article are arranged as follows. 
First, a full mathematical description of the magnetic diffusion problem is presented in section \ref{sec:problem-description}.
The derivation of an exact solution for the problem is provided in details in section \ref{sec:derive-analytical-solution}.
A test of the analytical solution by one-dimensional (1D) simulations is carried out in Section \ref{sec:test-analyical-solution}.
Finally, we conclude in section \ref{sec:conclusion}.

\section{Mathematical description of the magnetic diffusion problem}\label{sec:problem-description}

The so called ``magnetic diffusion problem'' is described by a magnetic diffusion equation and an Ohmic heating induced internal internal energy equation.
This paper concentrates on the problem's one-dimensional version, which is represented by the following two coupled equations:	
\begin{align}
\frac{\partial B(x,t)}{\partial t} &= \frac{\partial}{\partial x}\left( \frac{\eta(x,t)}{\mu_0} \frac{\partial B(x,t)}{\partial x} \right),\label{eq:mag-diffusion}\\
\frac{\partial e(x,t)}{\partial t} &= \eta(x,t) \left( \frac{1}{\mu_0} \frac{\partial B(x,t)}{\partial x} \right)^2,\label{eq:ohmic-heating}
\end{align}
where $B$ is the magnetic field, $e$ is the internal energy density (i. e., internal energy per volume) in the material, $\mu_0$ is the vacuum permeability constant, and $\eta$ is the resistivity in the material.
The term on the right-side of Eq. \eqref{eq:ohmic-heating} is named Ohmic heating, 
with $\frac{1}{\mu_0} \frac{\partial}{\partial x} B(x,t) \equiv j(x,t)$ corresponding to the electric current density in electromagnetics.

\begin{figure}
    \centering
    \includegraphics[width=0.5\linewidth]{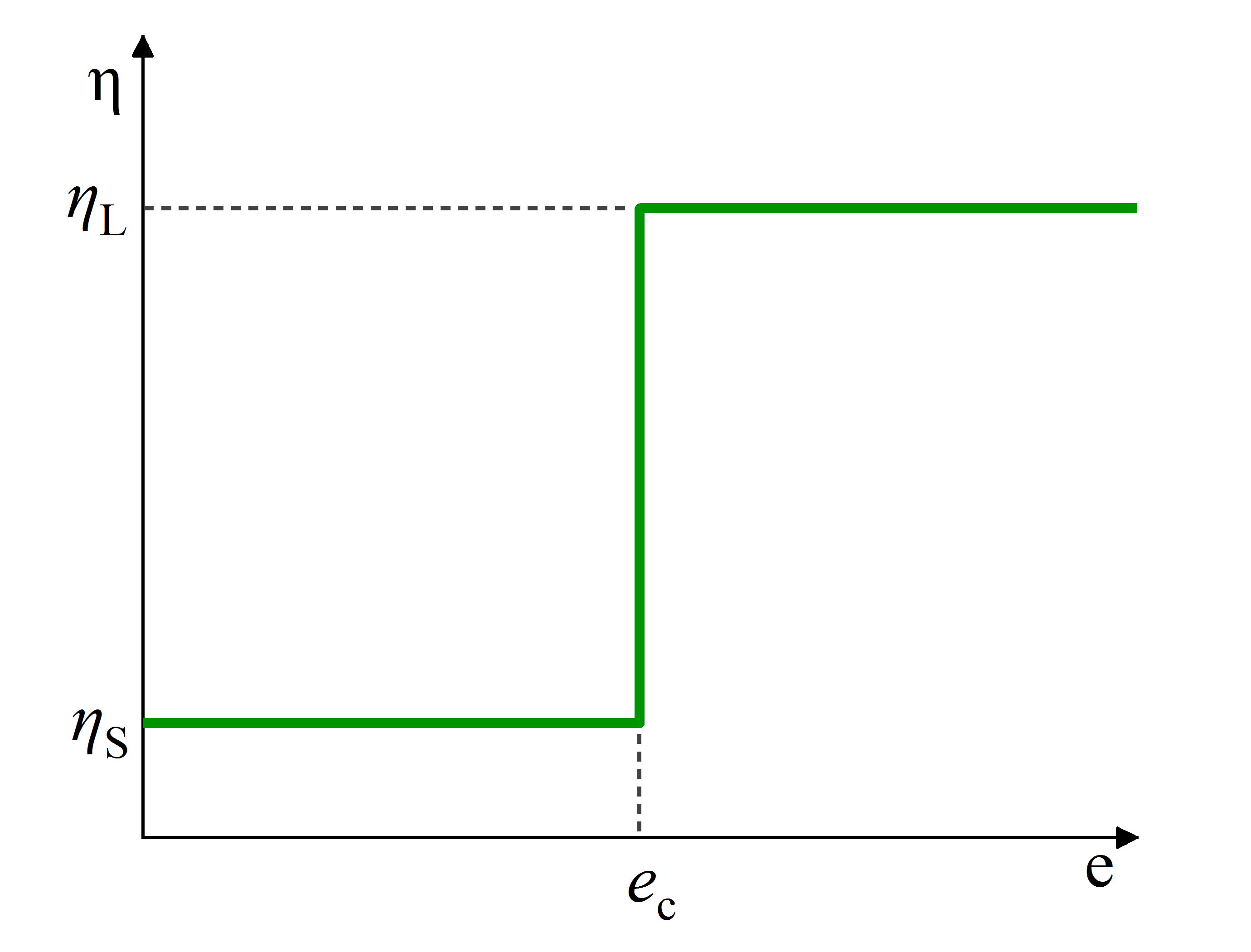}
    \caption{The step-function resistivity model.}
    \label{fig:step-function}
\end{figure}

The resistivity $\eta$ is considered to have a dependence on the internal energy density, that is,
\begin{equation}
\eta(x,t) = \eta(e)
\end{equation}
According to metals' resistivity model \cite{burgess1986},
the resistivity $\eta$ of a metal increases when the metal is heated from room temperature to higher,
and especially the resistivity has an abrupt up-jump when the temperature grows across the gasification point,
after that, the resistivity slides down gradually with the increase of temperature.
This resistivity model is simplified to a step-function in this paper, as described by Eq. \eqref{eq:step-function}
and illustrated in Fig. \ref{fig:step-function}.
\begin{equation}\label{eq:step-function}
\eta(e) = 
	\left\{
	\begin{array}{l}
		\eta_{\text{S}},\quad~~~~~~\text{for  }e <= e_{\text{c}} \\
		 \eta_{\text{L}},\quad~~~~~~\text{for  }e > e_{\text{c}} \\
	\end{array}
	\right.,
\end{equation}
where $\eta_{\text{L}}$, $\eta_\text{S}$, and $e_\text{c}$ are all constant parameters.

We consider the problem of a constant magnetic field in vacuum diffuses into a semi-infinite metal. It corresponds to the one-dimensional magnetic diffusion problem with the following boundary and initial conditions.
The magnetic field on the left boundary (i. e., $x = 0$) of the metal is a constant $B_0$, and the right boundary of the metal is located at infinity. 
At initial time, the magnetic field and internal energy density are both zero every where in the metal, that is, $B(x,t=0) = 0$ and $e(x,t=0) = 0$ for $x > 0$.
The main task in solving the problem is deriving the evolution function of magnetic field $B(x,t)$, 
while $e(x,t)$ is readily obtained through Eq. \eqref{eq:ohmic-heating} once $B(x,t)$ is known.

\section{Deriving an exact analytical solution for the problem}\label{sec:derive-analytical-solution}

\begin{figure}
    \centering
    \includegraphics[width=0.5\linewidth]{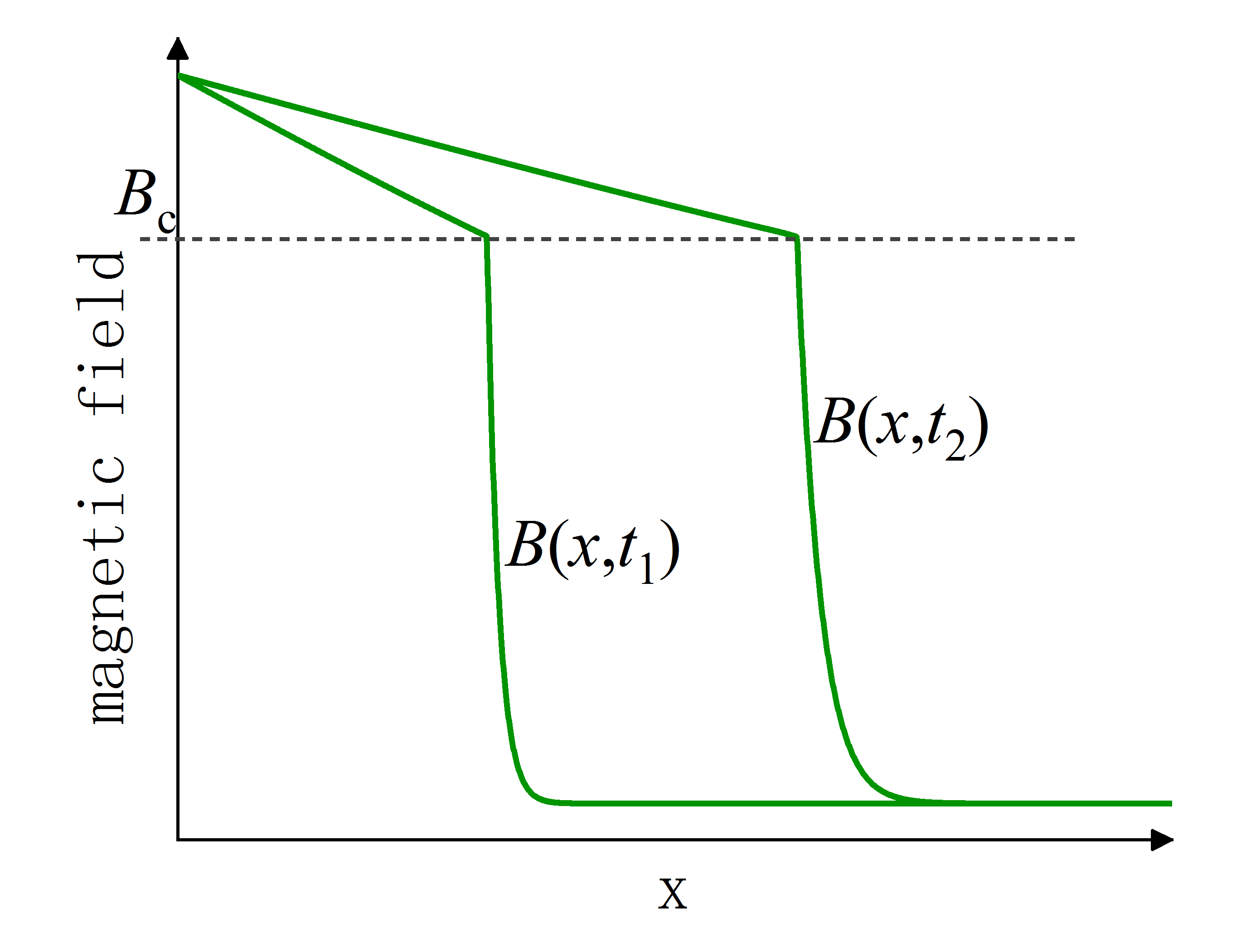}
    \caption{A guessing about the characteristics of the solution of $B(x,t)$.}
    \label{fig:solution-curve}
\end{figure}

We starts the derivation of solutions for the magnetic diffusion problem from a guessing: there exists a ``sharp-front'' solution of $B(x,t)$ that possesses self-similarity, as illustrated in Fig. \ref{fig:solution-curve}.
The phrase ``sharp-front'' means the curve of $B(x,t)$ along $x$ contains a ``knee'' that cuts the curve into two parts,
the part in front of (on the right side of) the knee corresponds to the ``cold'' region which possesses a small resistivity $\eta_\text{S}$, 
while that behind the knee corresponds to the ``burned'' region whose resistivity has jumped up to the larger value $\eta_\text{L}$.
According to the relation of $\eta$-$e$, it also means the internal energy density in front of the knee is below $e_\text{c}$,
while that back of it is above $e_\text{c}$.
For the self-similarity property of $B(x,t)$, it means for two distribution curves of magnetic field at two arbitrary times, $B(x,t_1)$ and $B(x, t_2)$,
they can completely coincide with each other if either of them is stretched or shrink along the $x$ direction properly.
While self-similarity is a well known property for the usual diffusion equation \cite{Zhu2005}, it is not already proved when the magnetic diffusion equation is coupled with Ohmic heating.
We are not able to provide a convincing proof for this guessing here, but if we are able to find in the following a solution for $B(x,t)$ based on the guessing, then its rationality is naturally confirmed.

\subsection{Transferring from B(x,t) to f(u)}

A direct conclusion from the self-similarity property is that the magnetic field on the knee is a time-independent constant, 
as the curve of $B(x,t)$ just stretches horizontally (i. e., along $x$) with time.
The constant magnetic field on the knee is tagged as $B_c$ in this paper.
Further more, from the self-similarity property, the function $B(x,t)$ can be written in the following form
\begin{equation}
B(x,t) = f(u),\quad~~~~~~\text{with }u=\frac{x}{x_c(t)},
\end{equation}
where, $x_c(t)$ is the position of the knee, and the fraction $x/x_c(t)$ is interpreted as a dimensionless scaled position that is represented by a symbol $u$.

The curve of $f(u)$ can be called ``a normalized curve of the distribution of $B(x,t)$ along $x$ at any time''.
The task of deriving solutions for $B(x,t)$ thus becomes solving $f(u)$.
The curve of $B(x,t)$ has a knee at $x=x_c$ tells that $f(u)$ has a knee at $u = 1$.
It is convenient to represent the two segments at left and right sides of the knee in curve of $f(u)$ by two different labels, that is,
\begin{equation}
f(u) = 
	\left\{
	\begin{array}{l}
		f_{\text{L}}(u),\quad~~~~~~\text{for  }u\in[0,1) \\
		f_{\text{S}}(u),\quad~~~~~~\text{for  }u\in[1,\infty) \\
	\end{array}
	\right.
\end{equation}
The functions $f_{\text{L}}(u)$ and $f_{\text{S}}(u)$ have the boundary conditions of 
\begin{equation}\label{eq:fl-boundary-condition}
	\left\{
	\begin{array}{l}
		f_{\text{L}}(0)=B_0 \\
		f_{\text{L}}(1)=B_c \\
	\end{array}
	\right.
\end{equation}
and 
\begin{equation}\label{eq:fs-boundary-condition}
	\left\{
	\begin{array}{l}
		f_{\text{S}}(1)=B_c \\
		f_{\text{S}}(\infty)=0 \\
	\end{array}
	\right.,
\end{equation}
respectively.

The function $f(u)$ is solved based on the equations \eqref{eq:mag-diffusion} and \eqref{eq:ohmic-heating}, and the boundary conditions \eqref{eq:fl-boundary-condition} and \eqref{eq:fs-boundary-condition}.
The partial differentiation of $B(x,t)$ in time and space can be written in $f(u)$ as
\begin{equation}\label{eq:diff-B-in-fu}
	\left\{
	\begin{array}{l}
		\frac{\partial B}{\partial t} = f^\prime(u) \frac{-x}{x_c^2}\frac{d x_c}{dt} \\
		\frac{\partial^2 B}{\partial x^2} = f^{\prime\prime}(u)\frac{1}{x_c^2} \\
	\end{array}
	\right.
\end{equation}
Inserting \eqref{eq:diff-B-in-fu} into the magnetic diffusion equation \eqref{eq:mag-diffusion}, we get the differential equation of $f(u)$ as
\begin{equation}\label{eq:mag-diff-eq-in-fu}
f^\prime(u) = -\frac{\eta}{\mu_0}\frac{1}{v_c x_c}\frac{1}{u}f^{\prime\prime}(u),
\end{equation}
where $f^\prime(u)$ and $f^{\prime\prime}(u)$ are the first-order and second-order derivatives of $f(u)$ to $u$, and $v_c \equiv dx_c/dt$ is the moving velocity of the knee.
Please note that Eq. \eqref{eq:mag-diff-eq-in-fu} holds separately in the $\eta_{\text{L}}$ and $\eta_{\text{S}}$ regions (i. e., $u<1$ and $u>1$),
and in deriving Eq. \eqref{eq:mag-diff-eq-in-fu} we have used the condition that
$\eta_{\text{L}}$ and $\eta_{\text{S}}$ are constants independent of $u$.
There is an important conclusion
\footnote{This conclusion is in fact just a different way of expression to the well known property for the self-similar solution of a usual diffusion equation, 
where it is usually stated like ``the self-similar variable is $x/\sqrt{t}$''.}
from Eq. \eqref{eq:mag-diff-eq-in-fu}, that is, the product of $(v_c x_c)$ is time-independent, 
since $u$ is a scaled dimensionless parameter which has no direct relation to time here.
So, we define a time-independent variable 
\begin{equation}\label{eq:h-define}
h \equiv \mu_0 v_c x_c
\end{equation}
to simplify the forms of equations.

Besides equation \eqref{eq:mag-diff-eq-in-fu}, there is an additional relation for $f(u)$ at the point of $u=1$ as
\begin{equation}\label{eq:mag-flux-continue-in-fu}
\left. \frac{\eta_{\text{L}}}{\mu_0}\frac{1}{x_c}f^\prime(u) \right|_{u=1^-} = \left. \frac{\eta_{\text{S}}}{\mu_0}\frac{1}{x_c}f^\prime(u) \right|_{u=1^+},
\end{equation}
which comes from the magnetic flux continuity condition of
\begin{equation}\label{eq:mag-flux-continue}
\left. \frac{\eta_{\text{L}}}{\mu_0}\frac{\partial B}{\partial x} \right|_{x=x_c^-} = \left. \frac{\eta_{\text{S}}}{\mu_0}\frac{\partial B}{\partial x} \right|_{x=x_c^+}
\end{equation}
Using the symbols $f_\text{L}$ and $f_\text{S}$, the equations \eqref{eq:mag-diff-eq-in-fu} and \eqref{eq:mag-flux-continue-in-fu} can be re-expressed more explicitly as
\begin{equation}\label{eq:mag-diff-eq-in-flfs}
	\left\{
	\begin{array}{l}
		f_{\text{L}}^\prime(u) = -\frac{\eta_{\text{L}}}{h}\frac{1}{u}f_{\text{L}}^{\prime\prime}(u) \\
		f_{\text{S}}^\prime(u) = -\frac{\eta_{\text{S}}}{h}\frac{1}{u}f_{\text{S}}^{\prime\prime}(u) \\
	\end{array}
	\right.
\end{equation}
and
\begin{equation}\label{eq:mag-flux-continue-in-flfs}
\eta_{\text{L}} f_{\text{L}}^\prime(1) = \eta_{\text{S}} f_{\text{S}}^\prime(1)
\end{equation}

The Ohmic heating energy equation \eqref{eq:ohmic-heating}, under the circumstance of the ``sharp-front'' solution, is transferred to the following condition:
for an \emph{arbitrary} position $x$, when the knee (whose position is represented by $x_c(t)$) has moved from 0 to this position, 
the Ohmic heating energy collected at this point should equal exactly to the critical energy density $e_c$.
According to Eq. \eqref{eq:ohmic-heating}, this condition is expressed mathematically as
\begin{equation}\label{eq:ec-condition}
e_c = \int_0^x\eta_{\text{S}}\left( \frac{1}{\mu_0} \frac{\partial B(x,t)}{\partial x} \right)^2\frac{d x_c}{v_c(x_c)}
\end{equation}
If expressed in function $f(u)$, it is
\begin{equation}\label{eq:ec-condition-in-fu}
e_c = \int_0^x\eta_{\text{S}}\left( \frac{1}{\mu_0} \frac{1}{x_c} f_{\text{S}}^\prime(x/x_c) \right)^2\frac{d x_c}{v_c(x_c)}
\end{equation}

Eqs. in \eqref{eq:mag-diff-eq-in-flfs}, together with the boundary conditions \eqref{eq:fl-boundary-condition} and \eqref{eq:fs-boundary-condition}, the connection condition \eqref{eq:mag-flux-continue-in-flfs}, and the integration condition \eqref{eq:ec-condition-in-fu},
constitute the complete conditions for solving the function $f(u)$.

\subsection{Solving f(u) and B(x,t)}\label{sec:analytical-solution}

From the differential equations in \eqref{eq:mag-diff-eq-in-flfs}, we get a general solution of $f^\prime(u)$ as
\begin{equation}\label{eq:fu-prime-solution}
	\left\{
	\begin{array}{l}
		f_{\text{L}}^\prime(u) = A_{\text{L}} \exp ( -\frac{1}{2}\frac{h}{\eta_{\text{L}}} u^2 ) \\
		f_{\text{S}}^\prime(u) = A_{\text{S}} \exp ( -\frac{1}{2}\frac{h}{\eta_{\text{S}}} u^2 ) \\
	\end{array}
	\right.,
\end{equation}
where, $A_{\text{L}}$ and $A_{\text{S}}$ are parameters remain to be determined.
Considering the solution \eqref{eq:fu-prime-solution} together with the connection condition \eqref{eq:mag-flux-continue-in-flfs},
$A_{\text{L}}$ and $A_{\text{S}}$ are then related as
\begin{equation}\label{eq:al-as-connection}
\frac{A_{\text{L}}}{A_{\text{S}}} = \frac{\eta_{\text{S}} \exp ( -\frac{1}{2}\frac{h}{\eta_{\text{S}}} )} {\eta_{\text{L}} \exp ( -\frac{1}{2}\frac{h}{\eta_{\text{L}}} )}
\end{equation}
Further more, the boundary conditions \eqref{eq:fl-boundary-condition} and \eqref{eq:fs-boundary-condition} can be re-expressed in $f^\prime(u)$ as
\begin{equation}\label{eq:fl-prime-integ}
\int_0^1 f^\prime_{\text{L}}(u) du = B_c - B_0
\end{equation}
and 
\begin{equation}\label{eq:fs-prime-integ}
\int_1^\infty f^\prime_{\text{S}}(u) du = -B_c
\end{equation}
Inserting \eqref{eq:fu-prime-solution} into them, we get
\begin{equation}\label{eq:al-solution}
A_{\text{L}} = \frac{B_c - B_0}{\int_0^1 \exp (-\frac{1}{2}\frac{h}{\eta_{\text{L}}} u^2) du}
\end{equation}
and
\begin{equation}\label{eq:as-solution}
A_{\text{S}} = \frac{-B_c}{\int_1^\infty \exp (-\frac{1}{2}\frac{h}{\eta_{\text{S}}} u^2) du}
\end{equation}
Then, the relation of \eqref{eq:al-as-connection} becomes
\begin{equation}\label{eq:bc-h-relation-one}
\frac{B_0-B_c}{B_c}\frac{\int_1^\infty \exp (-\frac{1}{2}\frac{h}{\eta_{\text{S}}} u^2) du}{\int_0^1 \exp (-\frac{1}{2}\frac{h}{\eta_{\text{L}}} u^2) du} = 
\frac{\eta_{\text{S}} \exp ( -\frac{1}{2}\frac{h}{\eta_{\text{S}}} )} {\eta_{\text{L}} \exp ( -\frac{1}{2}\frac{h}{\eta_{\text{L}}} )}
\end{equation}
Finally, inserting the expression of $f^\prime_{\text{S}}(u)$ in \eqref{eq:fu-prime-solution} into the integration condition \eqref{eq:ec-condition-in-fu},
we get
\footnote{
It can be seen from Eq. \eqref{eq:ec-condition-in-as} that the integration in Eq. \eqref{eq:ec-condition} is finally independent of $x$.
This is in fact a key for the existence of a self-similarity solution for the nonlinear magnetic diffusion problem, 
as a response to the discussion on the head of Sec. \ref{sec:derive-analytical-solution}.
}
\begin{equation}\label{eq:ec-condition-in-as}
e_c = \frac{1}{2\mu_0} \frac{\eta_{\text{S}}}{h} A_{\text{S}}^2 \int_0^1 \frac{1}{t}\exp(-\frac{h}{\eta_{\text{S}}}\frac{1}{t}) dt
\end{equation}
Taking the expression of $A_\text{S}$ in \eqref{eq:as-solution}, it becomes
\begin{equation}\label{eq:bc-h-relation-two}
e_c = \frac{B_c^2}{2\mu_0} \frac{\eta_{\text{S}} / h}{\left(\int_1^\infty \exp (-\frac{1}{2}\frac{h}{\eta_{\text{S}}} u^2) du \right)^2} \int_0^1 \frac{1}{t}\exp(-\frac{h}{\eta_{\text{S}}}\frac{1}{t}) dt
\end{equation}

Relations \eqref{eq:bc-h-relation-one} and \eqref{eq:bc-h-relation-two} form the group of equations for solving the two undetermined parameters $B_c$ and $h$.
If $B_c$ and $h$ are solved, $A_{\text{L}}$ and $A_{\text{S}}$ are then determined through \eqref{eq:al-solution} and \eqref{eq:as-solution}
, and the first derivative of $f(u)$ is then determined through \eqref{eq:fu-prime-solution}.
Finally, the function $f(u)$ is obtained by integrating $f^\prime(u)$.

The position of the moving knee, $x_c(t)$, is determined according to the relation \eqref{eq:h-define}, through which we get
\begin{equation}\label{eq:dxc-dt}
\frac{dx_c}{dt} = \frac{h}{\mu_0}\frac{1}{x_c},
\end{equation}
whose solution is
\begin{equation}\label{eq:xc-t-solution}
x_c(t) = \sqrt{(2h/\mu_0)t}
\end{equation}
With $x_c(t)$ and the function $f(u)$, $B(x,t)$ is finally determined through
\begin{equation}\label{eq:b-x-t-solution}
B(x,t) = f(x/x_c(t))
\end{equation}

\subsection{Calculating Bc and h by numerical integration}

Even though $B(x,t)$ is formally solved in subsection \ref{sec:analytical-solution}, there are constant quantities $B_c$ and $h$ remain
to be determined according to equations \eqref{eq:bc-h-relation-one} and \eqref{eq:bc-h-relation-two}.
Those equations are too complex to be resolved analytically, and here we discuss a numerical way for calculating $B_c$ and $h$.
In principle, $B_c$ and $h$ are functions of the four input parameters $e_c$, $\eta_{\text{L}}$, $\eta_{\text{S}}$, and $B_0$, 
that is,
\begin{equation}\label{eq:bc-h-in-input-parameters}
    \left\{
    \begin{array}{l}
        B_c = B_c(e_c, \eta_{\text{L}}, \eta_{\text{S}}, B_0) \\
        h = h(e_c, \eta_{\text{L}}, \eta_{\text{S}}, B_0) \\
    \end{array}
    \right.
\end{equation}
The task is thus to figure out numerical values for $B_c$ and $h$, 
when the values of the four parameters $e_c$, $\eta_{\text{L}}$, $\eta_{\text{S}}$, and $B_0$ are given.

The numerical way goes as follows. 
When the values of the four input parameters $e_c$, $\eta_{\text{L}}$, $\eta_{\text{S}}$, and $B_0$ are given,
Eq. \eqref{eq:bc-h-relation-one} can be viewed as a function of $B_c$ to $h$, labelled as $B_c^{(1)}(h)$,
and so does Eq. \eqref{eq:bc-h-relation-two}, labelled as $B_c^{(2)}(h)$. These two functions can be written explicitly as
\begin{equation}\label{eq:bc-to-h-one}
B_c^{(1)}(h) = B_0
\left( 1+\frac{\eta_{\text{S}}}{\eta_{\text{L}}}
\frac{\exp ( -\frac{1}{2}\frac{h}{\eta_{\text{S}}} )}{\exp ( -\frac{1}{2}\frac{h}{\eta_{\text{L}}} )}
\frac{\int_0^1 \exp (-\frac{1}{2}\frac{h}{\eta_{\text{L}}} u^2) du}{\int_0^1 \exp (-\frac{1}{2}\frac{h}{\eta_{\text{S}}} \frac{1}{t^2}) \frac{1}{t^2} dt}
\right) ^{-1}
\end{equation}
and
\begin{equation}\label{eq:bc-to-h-two}
B_c^{(2)}(h) = \sqrt{2\mu_0 e_c} 
\left(  
\frac{h}{\eta_{\text{S}}}
\frac{ \left(\int_0^1 \exp (-\frac{1}{2}\frac{h}{\eta_{\text{S}}} \frac{1}{t^2}) \frac{1}{t^2} dt \right)^2 }{\int_0^1 \frac{1}{t}\exp(-\frac{h}{\eta_{\text{S}}}\frac{1}{t}) dt}
\right)^\frac{1}{2},
\end{equation}
where, $\int_0^1$ integration's have been used to replace $\int_1^\infty$ integration's by the $du\to d(1/t)$ replacement, for the convenience of numerical integration.
The curves of $B_c^{(1)}(h)$ and $B_c^{(2)}(h)$ can be drawn by doing integration's numerically in Eqs. \eqref{eq:bc-to-h-one} and \eqref{eq:bc-to-h-two} under different values of $h$, and then the intersection point of the two curves, which forms the solution for $B_c$ and $h$, is picked out numerically.

However, it is worth noting that, the types of integration's 
\[
I_1(a) = \int_0^1\exp(-a \frac{1}{t^2})\frac{1}{t^2} dt,
\]
\[
I_2(a) = \int_0^1\exp(-a \frac{1}{t})\frac{1}{t} dt,
\]
and
\[
I_3(a) = \int_0^1\exp(-au^2)du
\]
are not numerically stable when $a$ is large.
This issue can be cured by the transformations of
\begin{equation}\label{eq:integration-I1}
\Tilde{I}_1(a) \equiv \exp(a) I_1(a) = \int_0^1\exp\left(a(1-\frac{1}{t^2})\right)\frac{1}{t^2} dt,
\end{equation}
\begin{equation}\label{eq:integration-I2}
\Tilde{I}_2(a) \equiv \exp(a) I_2(a) = \int_0^1\exp\left(a(1-\frac{1}{t})\right)\frac{1}{t} dt,
\end{equation}
and
\begin{equation}\label{eq:integration-I3}
\Tilde{I}_3(a) \equiv \exp(a) \int_0^1\exp(-au^2)du = \int_0^1\exp\left(a(1-u^2)\right)du
\end{equation}
With those integration functions defined in \eqref{eq:integration-I1} - \eqref{eq:integration-I3}, $B_c^{(1)}(h)$ and $B_c^{(2)}(h)$ can be expressed as
\begin{equation}\label{eq:bc-to-h-one-in-I}
B_c^{(1)}(h) = B_0
\left( 
1 + \frac{\eta_{\text{S}}}{\eta_{\text{L}}}  \frac{\Tilde{I}_3(\frac{1}{2}\frac{h}{\eta_{\text{L}}})}{\Tilde{I}_1(\frac{1}{2}\frac{h}{\eta_{\text{S}}})}
\right) ^{-1}
\end{equation}
and
\begin{equation}\label{eq:bc-to-h-two-in-I}
B_c^{(2)}(h) = \sqrt{2\mu_0 e_c} 
\left(  
\frac{h}{\eta_{\text{S}}}
\frac{ \Tilde{I}_1^2(\frac{1}{2}\frac{h}{\eta_{\text{S}}}) }{ \Tilde{I}_2(\frac{h}{\eta_{\text{S}}}) }
\right)^\frac{1}{2}
\end{equation}

\subsection{Some more discussions about Bc and h}\label{sec:discuss-bc-h}

The above subsection has provided enough information for calculating $B_c$ and $h$ numerically.
However, since the physical meanings of the two quantities $B_c$ and $h$ are clear and important:
$B_c$ is the magnetic field on the knee of the sharp-front magnetic distribution, 
and $h$ represents the penetrating velocity of the diffusing magnetic field,
it is beneficial to gain more analytical knowledge about them.

By taking $B_c^{(1)}(h) = B_c^{(2)}(h)$ in Eqs. \eqref{eq:bc-to-h-one-in-I} and \eqref{eq:bc-to-h-two-in-I}, we get
the equation of $h$ as
\begin{equation}\label{eq:h-equation}
\left( 
1 + \frac{\eta_{\text{S}}}{\eta_{\text{L}}}  \frac{\Tilde{I}_3(\frac{1}{2}\frac{h}{\eta_{\text{L}}})}{\Tilde{I}_1(\frac{1}{2}\frac{h}{\eta_{\text{S}}})}
\right)
\left(  
\frac{h}{\eta_{\text{S}}}
\frac{ \Tilde{I}_1^2(\frac{1}{2}\frac{h}{\eta_{\text{S}}}) }{ \Tilde{I}_2(\frac{h}{\eta_{\text{S}}}) }
\right)^\frac{1}{2}
=
\frac{B_0}{\sqrt{2\mu_0 e_c}}
\end{equation}
From \eqref{eq:h-equation}, we get to know two things about $h$. First, the dependence of $f$ on $B_0$ and $e_c$
is simplified to the dependence on a single value $b \equiv B_0 / \sqrt{2\mu_0 e_c}$. Second, if we define a ratio $r \equiv \eta_{\text{L}} / \eta_{\text{S}}$,
the left side of \eqref{eq:h-equation} can be written as
\[
\left( 
1 + \frac{1}{r}  \frac{\Tilde{I}_3(\frac{1}{2}\frac{h}{\eta_{\text{L}}})}{\Tilde{I}_1(\frac{1}{2}\frac{h}{\eta_{\text{L}}}r)}
\right)
\left(  
\frac{h}{\eta_{\text{L}}}r
\frac{ \Tilde{I}_1^2(\frac{1}{2}\frac{h}{\eta_{\text{L}}}r) }{ \Tilde{I}_2(\frac{h}{\eta_{\text{L}}}r) }
\right)^\frac{1}{2}
\]
It means $h$ is proportional to $\eta_\text{L}$ (or $\eta_\text{S}$) under a given ratio $r$.
Combining those two things, the dependence of $h$ on the four input parameters can be simplified to the form of
\begin{equation}\label{eq:h-expression-simplified}
h(B_0, e_c, \eta_\text{L}, \eta_\text{S}) = \eta_\text{L} \mathscr{H}(b, r)
\end{equation}
Inserting \eqref{eq:h-expression-simplified} into \eqref{eq:bc-to-h-one-in-I}, we get an expression for $B_c$ as
\[
B_c(B_0, e_c, \eta_\text{L}, \eta_\text{S}) = B_0
\left( 
1 + \frac{1}{r}  \frac{\Tilde{I}_3(\frac{1}{2}\mathscr{H})}{\Tilde{I}_1(\frac{1}{2}\mathscr{H}r)},
\right) ^{-1}
\]
which can be written in an abbreviated form as
\begin{equation}\label{eq:bc-expression-simplified}
B_c(B_0, e_c, \eta_\text{L}, \eta_\text{S}) = B_0 \mathscr{B}(b, r)
\end{equation}
 
\section{Test the exact solution}\label{sec:test-analyical-solution}

The exact analytical solution derived in the above section is verified by comparing it to the 1D simulation, that is:
for a given set of input values of the four parameters ($B_0$, $e_c$, $\eta_{\text{L}}$, $\eta_{\text{S}}$),
the 1D simulation obtains the distribution of $B(x,t)$ by evolving the magnetic diffusion problem from its initial condition to the time $t$,
while the analytical solution puts out $B(x,t)$ directly, 
the two $B(x,t)$'s are then compared as a verification of the analytical solution.
It is known that the result of 1D simulation would inevitably have some departure from the theoretical ideal solution, 
and by refining the mesh in 1D simulation, the result should approach the ideal one. 
In the following, a set of values $B_0 = 0.2$, $e_c = 0.1$, $\eta_{\text{L}} = 9.7\times 10^{-3}$, and $\eta_{\text{S}} = 9.7\times 10^{-5}$ are adopted, as an example for testing the analytical solution. (The group of physical units ``cm'', ``us'', ``$10^3$Tesla'', ``$10^5\text{J}/\text{cm}^3$'', and ``$10^2\text{m}\Omega\text{.cm}$'', for time, length, magnetic field, energy density, and resistivity, is assumed being applied all through this article.)

\begin{figure}
    \centering
    \includegraphics[width=0.5\linewidth]{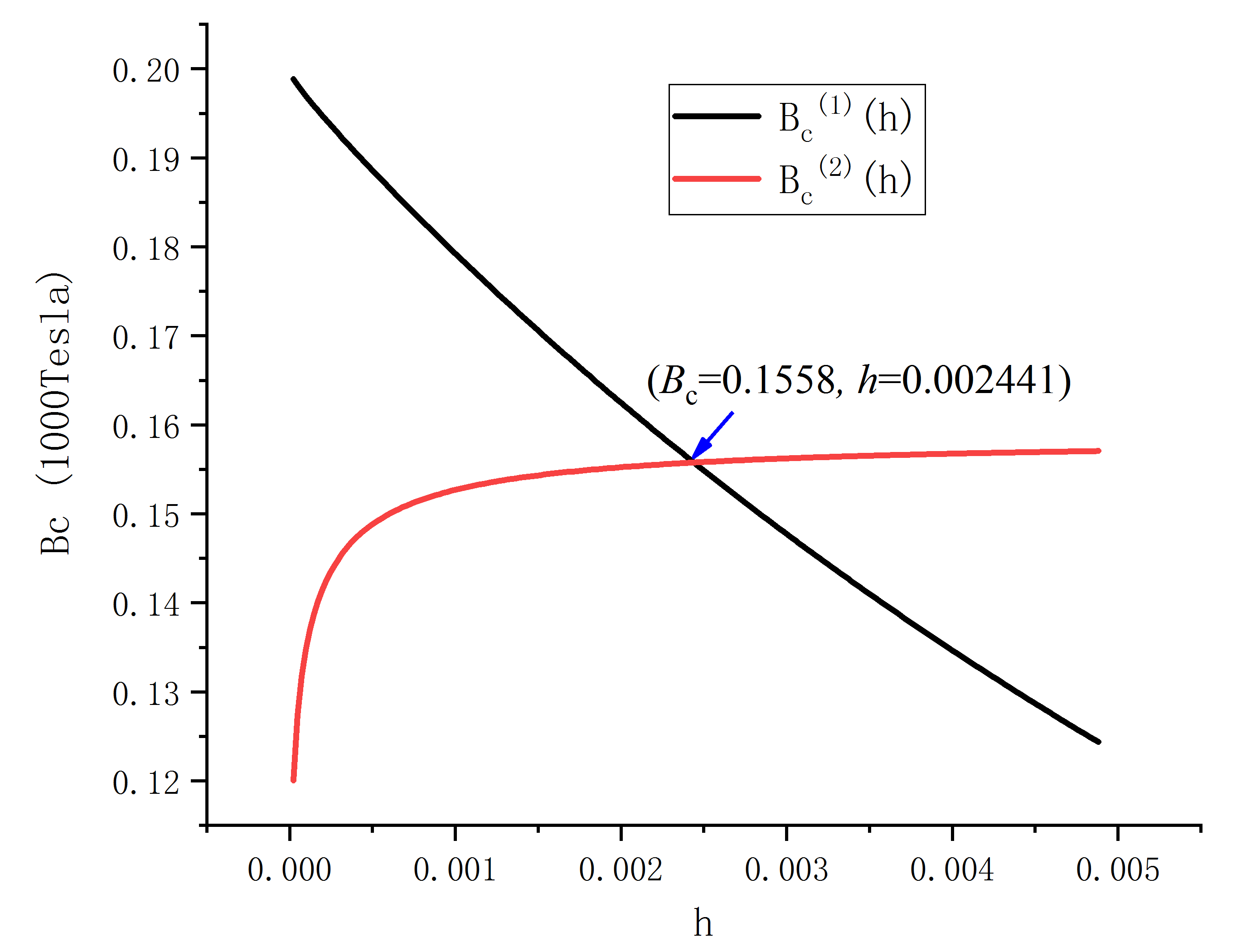}
    \caption{Intersection of the two curves of $B_c^{(1)}(h)$ and $B_c^{(2)}(h)$.}
    \label{fig:bc-h-intersection}
\end{figure}

\begin{figure}
    \centering
    \includegraphics[width=0.5\linewidth]{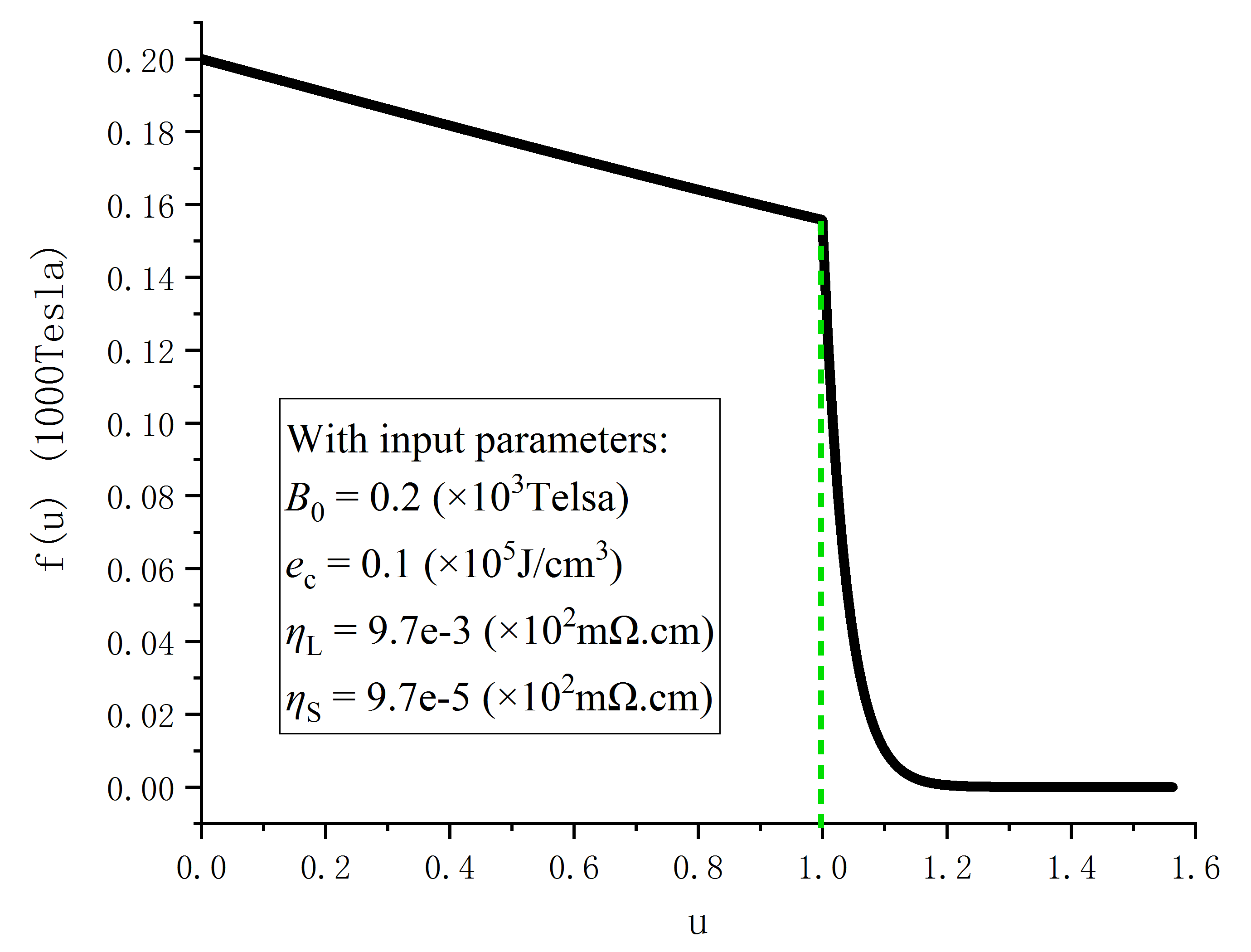}
    \caption{The curve of $f(u)$ under the given set of input parameters.}
    \label{fig:f-u-curve}
\end{figure}

\begin{figure}
    \centering
    \includegraphics[width=0.45\linewidth]{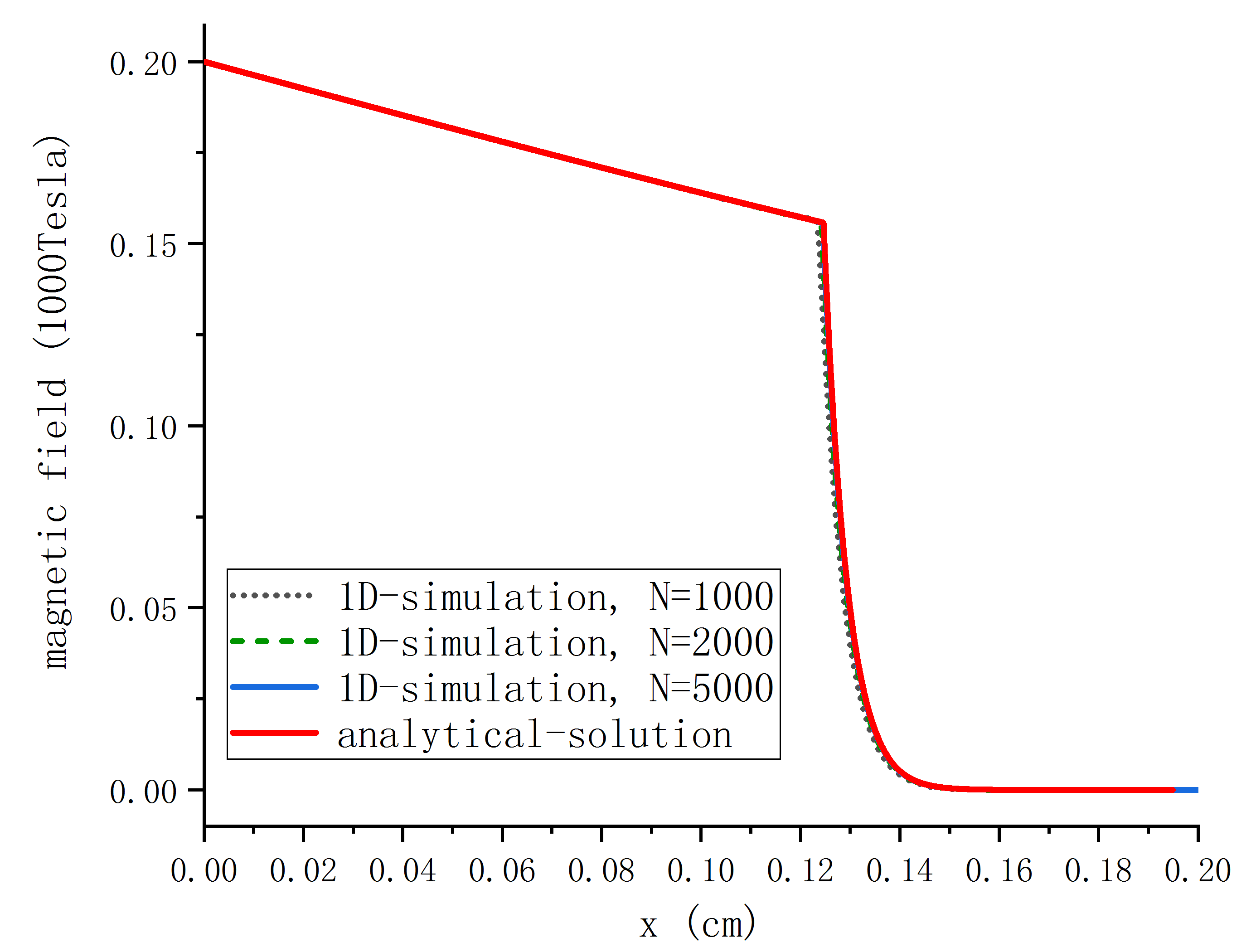}
    \includegraphics[width=0.45\linewidth]{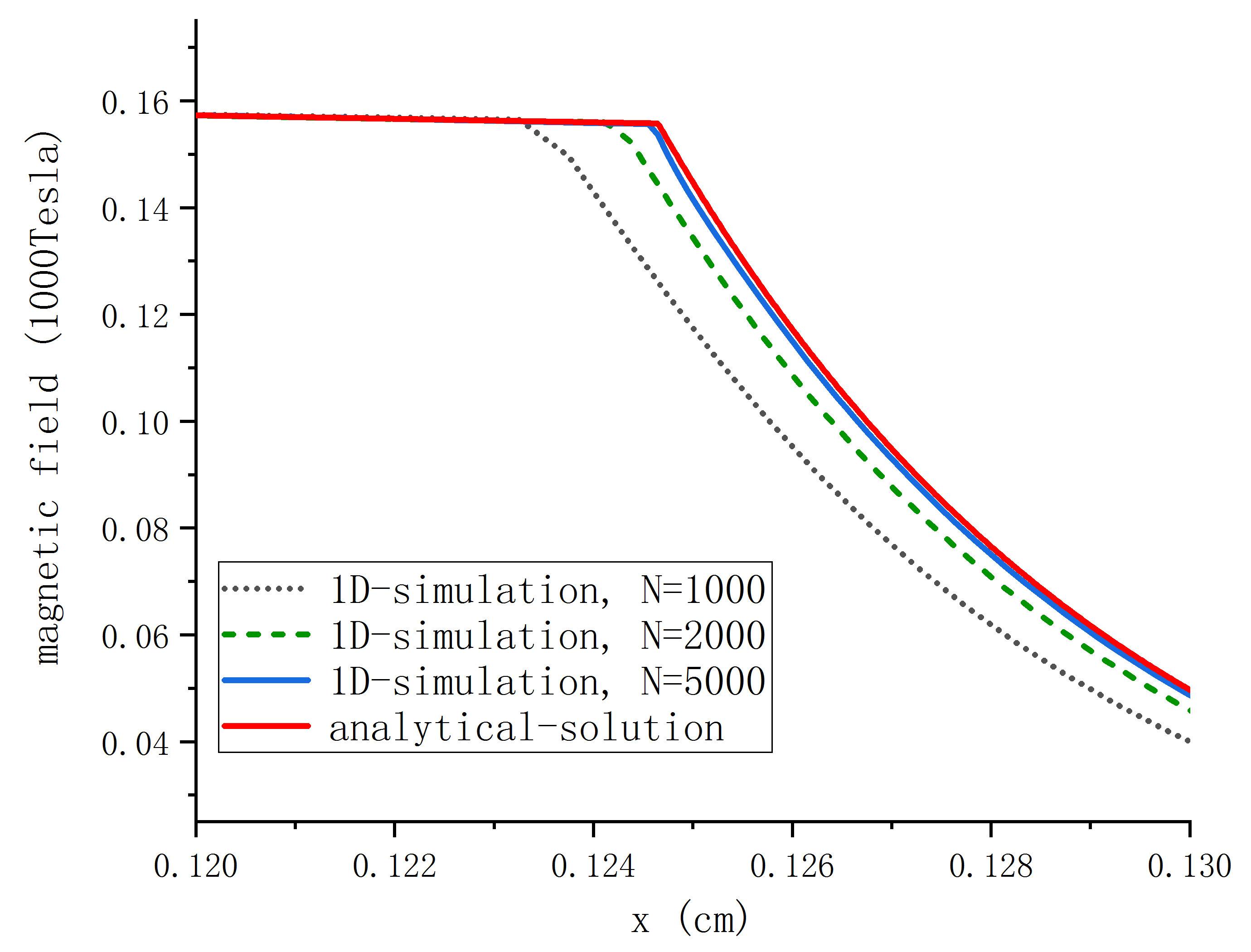}
    \caption{Comparison of the 1D simulated and the analytical solutions for $B(x,t)$. The quantity $N$ is the amount of grids used in 1D simulations in a space range of $0 < x < 0.5$. (The right figure is an enlarged view of the left one around the knee.)}
    \label{fig:b-x-t-compare}
\end{figure}

We use a numerical scheme from \cite{yan2022} for the 1D simulation.
The 1D simulation is run to $t=0.4$. Different mesh sizes are used for the simulation, the resulted $B(x,t)$'s are presented in Fig. \ref{fig:b-x-t-compare}.
The analytical solution of $B(x,t)$ is obtained following the steps in Sec. \ref{sec:derive-analytical-solution}.
First, the $B_c$ and $h$ is figured out by finding the intersection of the two curves $B_c^{(1)}(h)$ and $B_c^{(2)}(h)$,
the result is shown in Fig. \ref{fig:bc-h-intersection}, where $B_c = 0.1558$ and $h = 2.441\times 10^{-3}$.
Second, the curve of $f(u)$ is obtained through integrating $f^\prime(u)$ which is given by Eq. \eqref{eq:fu-prime-solution}.
To make the numerical integration of $f^\prime(u)$ convenient, we have rewritten \eqref{eq:fu-prime-solution}, by using the integration functions
defined in \eqref{eq:integration-I1} - \eqref{eq:integration-I3}, into the form of
\begin{equation}\label{eq:fu-prime-solution-in-I}
	\left\{
	\begin{array}{l}
		f_{\text{L}}^\prime(u) = \frac{B_c-B_0}{I_3(\frac{1}{2}\frac{h}{\eta_{\text{L}}})} \exp(\frac{1}{2} \frac{h}{\eta_{\text{L}}}(1-u^2)) \\
		f_{\text{S}}^\prime(u) = \frac{-B_c}{I_1(\frac{1}{2}\frac{h}{\eta_{\text{S}}})} \exp(\frac{1}{2} \frac{h}{\eta_{\text{S}}}(1-u^2)) \\
	\end{array}
	\right.
\end{equation}
The curve of $f(u)$ obtained is shown in Fig. \ref{fig:f-u-curve}.
Finally, the function $B(x,t)$ is obtained by substituting the $u$ in $f(u)$ by $x/x_c(t)$, where $x_c(t)$ is given by \eqref{eq:xc-t-solution}.
For drawing the curve of $B(x,t)$, it is simply achieved by re-scaling the horizontal coordinate of the curve of $f(u)$ with $1\to x_c(t)$, and the result is presented in Fig. \ref{fig:b-x-t-compare}.
From the comparison of the 1D simulated and the analytical solutions for $B(x,t)$ in Fig. \ref{fig:b-x-t-compare}, 
we can see the 1D simulation result approaches the analytical solution when a finer-and-finer mesh is used, 
and so, the correctness of the analytical solution is confirmed.

\section{Conclusions and discussions}\label{sec:conclusion}

A ``sharp-front'' type exact analytical solution for the magnetic diffusion problem is derived in this article, which is a solution with self-similarity.
There are two important time-independent quantities about the solution, $B_c$ and $h$, which represent the magnetic field on the knee and
the penetrating velocity of the diffusing magnetic field, respectively.
With given ratio's of $B_0 / \sqrt{2\mu_0 e_c}$ and $\eta_\text{L} / \eta_\text{S}$, the magnetic field on knee is proportional to the boundary magnetic field $B_0$,
and the penetrating velocity is proportional to $\eta_\text{L}$.
The analytical solution is confirmed by 1D simulations.
As the analytical solution that can be found for the nonlinear magnetic diffusion problem is seldom,
the finding of such an exact solution is valuable in mathematics.
What's more, the exact solution could also serve as a valuable benchmark example for testing simulation methods for the magnetic diffusion problem.

A copy of the computer code for calculating the exact solution following the steps described in Sec. \ref{sec:test-analyical-solution} can be found in \url{https://to-be-determined}.

\bibliographystyle{alpha}
\bibliography{sample}

\end{document}